\documentclass{elsart}
\usepackage{natbib}
\usepackage{graphicx}
\usepackage{amssymb}
\usepackage{amsmath}
\usepackage{bm}

\def\url#1{{\ttfamily\def\/{/\discretionary{}{}{}}#1}}
\def\bibcode#1{}

\begin{document}
\begin{frontmatter}
\title{Scale-dependent bias and the halo model}
\author{A.E. Schulz},
\author{Martin White}
\address{Departments of Physics and Astronomy,\\
University of California, Berkeley, CA 94720}
\thanks[aesemail]{schulz@astro.berkeley.edu}
\thanks[mjwemail]{mwhite@berkeley.edu}

\begin{abstract}
We use a simplified version of the halo model with a power law power spectrum
to study scale dependence in galaxy bias at the very large scales relevant to
baryon oscillations.   In addition to providing a useful pedagogical
explanation of the scale dependence of galaxy bias, the model provides
an analytic tool for studying how changes in the Halo Occupation Distribution
(HOD) impact the scale dependence of galaxy bias on scales
$10<k^{-1}<1000\,h^{-1}$Mpc, which is useful for interpreting the results of
complex N-body simulations.
We find that changing the mean number of galaxies per halo of a given mass
will change the scale dependence of the bias, but that changing the way the
galaxies are distributed within the halo has a smaller effect on the scale
dependence of bias at large scales.  We use the model to explain the
decay in amplitude of the baryon oscillations as $k$ increases, and
generalize the model to make predictions about scale dependent galaxy
bias when redshift space distortions are introduced. 
\end{abstract}
\end{frontmatter}

\section{Introduction}

The coupling of photons and baryons by Thompson scattering in the early 
universe results in gravity driven acoustic oscillations of the photon-baryon
fluid.  
The features that appear in both the Cosmic Microwave Background (CMB) 
anisotropies and matter power spectra are snapshots of the phases of these
oscillations at the time of decoupling, and provide important clues 
used to constrain a host of cosmological parameters.  Features in 
the matter power spectrum, referred to as baryon (acoustic)
oscillations, have the potential to strongly constrain the expansion
history of the universe and the nature of the dark energy.
These features in the matter power spectrum induce correlations in the
large-scale clustering of the IGM, clusters or galaxies.
Indeed a large, high redshift, spectroscopic galaxy survey encompassing a
million objects has been proposed \cite{WhitePaper} as a dark energy probe,
building on the successful detection (at low redshift) of the features in
the clustering of the luminous red galaxy sample of the Sloan Digital Sky
Survey \cite{SDSS}.
Key to this program is the means to localize the primordial features in
the galaxy power spectra, which necessarily involves theoretical understanding
of such complex issues as galaxy bias, non-linear structure evolution, and
redshift space distortions.  These effects can shift the observed 
scales of peaks and troughs in observations of the baryon oscillations,
affecting the transverse and radial measurements differently.  Marginalizing
over this uncertain scale dependence can be a large piece of the error budget
of proposed surveys.

Inroads into modeling galaxy bias, non-linear evolution, and redshift 
space distortions have been made by several groups using N-body
simulations \cite{Sims}, but these simulations are complex and often
mask the underlying physical mechanisms.  A route to understanding
the sophisticated simulations is provided by an analytic tool known as
the halo model \cite{Halo}.
The halo model makes the assumption that all matter in the universe lives
in virialized halos, and that the power spectrum of this matter can be
divided up into two distinct contributions; the 1-halo term arising from
the correlation of matter with other matter existing in the same halo,
and the 2-halo term quantifying the correlation of matter with matter
that lives in a different halo.
The halo model is extended to calculate galaxy power spectra through the
introduction of a Halo Occupation Distribution (HOD), that describes the
number of galaxies that exist in a halo of a given mass, and their
distribution inside that halo.
The HOD will naturally depend on the details of galaxy formation, but is
not in itself a theory of galaxy formation.  Rather it is a description of
the effects of galaxy formation on the number and distribution of the
galaxies.  The impact of galaxy formation on cosmological observables such
as the baryon oscillation can be studied in the halo model by investigating
the sensitivity of the observable to changes made to the HOD.

In this paper, we employ a simple version of the analytic halo model
to study the origins of scale dependence in galaxy bias.
We investigate the impact of changing the HOD on the observed spectra,
and show that the scale dependence of the bias arises in a natural way from
extending the dark matter description to a description of galaxies.
Specifically, in generalizing the description to an ensemble of rare
tracers of the dark matter, the 1-halo and 2-halo terms in the power
spectrum are shifted by different amounts.
The scale dependence in the galaxy bias arises from this difference.
We find that for small $k$, the galaxy power spectrum is sensitive to
the number of galaxies that occupy a halo, but not to their positions
within the halo.  We quantify the impact of redshift space distortions,
which find a natural description in the halo model, and discuss the
implications for large galaxy redshift surveys.

\section{The Halo Model}

We will try to understand the scale-dependence of the galaxy bias by
examining a simple model.  Our goal is not to make precise predictions,
but rather to use simple analytic approximations to help interpret the
results of N-body simulations, which are much more appropriate for studying
the fine details.

Our investigation makes use of the halo model \cite{Halo}, which assumes that
all of the mass and galaxies in the universe live in virialized halos,
whose clustering and number density is characterized by their mass.
This model can be used to approximate the two point correlation function of
the mass, and of various biased traces of the mass, such as luminous galaxies.
In this framework, there are two contributions to the power spectrum:
one arises from the correlation of objects that reside in the same halo
(the 1-halo term), and the other comes from the correlation of objects that
live in separate halos (the 2-halo term).  For the dark matter, for example,
the dimensionless power spectrum can be written
\begin{align}
  \Delta^2_{\rm dm} \equiv \frac{k^3 \, P_{\rm dm}(k)}{2\pi^2}
  \, =\, {}_{\rm 1h}\Delta^2_{\rm dm} + \,_{\rm 2h}\Delta^2_{\rm dm}
\end{align}
where \cite{Halo}:
\begin{align}
\label{dm1h}
  {}_{\rm 1h}\Delta^2_{\rm dm}&=\frac{k^3}{2 \pi^2}\, \frac{1}{\bar{\rho}^2}\,
  \int_{0}^\infty dM  \, n_h(M) M^2\, |y(M,k)|^2
\\
\label{dm2h}
  {}_{\rm 2h}\Delta^2_{\rm dm}&=\Delta^2_{\rm lin}\left[\frac{1}{\bar{\rho}}\,
  \int_{0}^\infty dM \, \,n_h(M) \, b_h(M,k) \, M \, y(M,k) \right]^2
\end{align}
with $M$ the virial mass of the halo, $\bar{\rho}$ the mean background 
density, $n_h(M)$ the number density of halos of a given virial mass,
and $b_h(M,k)$ the halo bias \cite{PeaksBias}.
The function $y(M,k)$ is the Fourier transform of the halo profile which
describes how the dark matter is spatially distributed within the halo.

Expressed this way, Eqs.~(\ref{dm1h}) and (\ref{dm2h}) lend themselves to a
fairly intuitive interpretation.
In the two halo term, the dark matter being correlated lives in widely
separated halos, of different masses.
For each of the two halos, the mass is multiplied by the function $y$ that
governs its spatial distribution within a halo, weighted by the number 
density of halos $n_h$ with bias $b_h$, and integrated over all possible 
halo masses.  The one halo term is even simpler -- correlating two bits of
dark matter residing in the same halo.  Thus there are two factors of
$M\times y$ weighted with the number density of halos $n_h$, and integrated
over all halo masses.   

We can generalize this framework to compute the 1- and 2-halo terms for
a galaxy population that traces the density field \cite{Halo}.
We divide the galaxies into two sub-populations, centrals and satellites.
The centrals will reside at the center of the host halo, while the
satellites will trace the dark matter.
The Halo Occupation Distribution (HOD) sets the number of tracers in a 
halo of mass M.  We assume there is either a central galaxy or not, and
the number of satellites is Poisson distributed \cite{KBWKGAP}.
For our model we will use 
\begin{align}
  \left\langle N_c \right\rangle &= \Theta(M-M_{\rm min}) \\
  \left\langle N_s \right\rangle &= \Theta(M-M_{\rm min}) 
    \left(\frac{M}{M_{\rm sat}}\right)^a
\end{align}
where $\Theta$ is the Heaviside function, and $M_{\rm min} < M_{\rm sat}$.
Note that the central galaxies do not trace the halo profile, and are not
weighted by $y$. 
The generalization of the 1-halo and 2-halo terms is given by \cite{Halo}
\begin{align}
  _{\rm 2h}\Delta^2_{\rm g}&=\Delta^2_{\rm lin}
  \left[ \frac{1}{\bar{n}_g} \int_{M_{\rm min}}^\infty dM \,
  n_h(M) \, b_h(M) \,
  \left(1+\left( \frac{M}{M_{\rm sat}}\right)^a y(M,k) \right) \right]^2 \\
  _{\rm 1h}\Delta^2_g&=\frac{k^3}{2 \pi^2}\frac{1}{\bar{n}_g^2} 
  \int_{M_{\rm min}}^\infty dM\, n_h(M) \,
    \left( 2 \left( \frac{M}{M_{\rm sat}}\right)^a y(M,k)
  + \left( \frac{M}{M_{\rm sat}}\right)^{2 a} |y(M,k)|^2 \right)
\end{align}
where
\begin{align}
  \bar{n}_{\rm gal}=\int_{M_{\rm min}}^\infty dM \,n_h(M) \, \left( 1+
  \left(\frac{M}{M_{\rm sat}}\right)^a \right)
\end{align}
is the number density of galaxies.
The interpretation of these expressions is similar to that of the dark matter.

For the purposes of this toy model, we shall adopt a power law for the
linear power spectrum,
\begin{align}
  \Delta^2_{\rm lin}=\left(\frac{k}{k_\star}\right)^{3+n}=\kappa^{3+n}
\end{align}
and define a dimensionless wavenumber $\kappa\equiv k/k_\star$.
In order to simplify many of the expressions in the calculation, we find 
it useful to change variables from $M$ to a dimensionless quantity, $\nu$, 
related to the peak height of the overdensity. For the power law model
considered here $\nu$ is a simple function of the mass:
\begin{align}\label{nudef}
  \nu(M)=\left(\frac{\delta_c}{\sigma(M)}\right)^2 =
         \left(\frac{M}{M_\star}\right)^{(n+3)/3} = m^{(n+3)/3} \quad .
\end{align}
Here, $\delta_c=1.686$ and $\sigma(M)$ is the linear theory variance 
in top hat spheres of radius $R=(3M/4\pi\bar{\rho})^{1/3}$.
We have introduced a a dimensionless mass $m$ in terms of the scale
mass $M_\star$, the mass for which $\sigma(M_\star)=\delta_c$ and $\nu=1$.  
Note $M_\star$ is a function of the power spectrum normalization $k_\star$
and the index $n$, and can be computed using the relation
\begin{align} \label{eqn:sigdef}
  \sigma^2(R)=\int_0^\infty \frac{dk}{k}\ \Delta^2_{\rm lin} \left[
  \frac{3j_1(kR)}{kR}\right]^2
\end{align}
where $j_1$ is the spherical Bessel function of order 1. 

The mass function $n_h(M)$ takes a simple form when expressed in terms of
the multiplicity function, $f(\nu)$.  The multiplicity function is a
normalized number density of halos at a given mass:
\begin{align}
  f(\nu) d\nu = \frac{M}{\bar{\rho}}\,n_h(M)\,dM
  \qquad {\rm with} \qquad
  \int f(\nu) \, d\nu=1 \quad .
\end{align}
For the Press-Schechter (P-S) mass function \cite{PreSch}
\begin{align}
  f(\nu) = \frac{e^{-\nu/2}}{\sqrt{2\pi\nu}}  
\end{align}
and the halos form biased tracers of the density field.
On large scales, for small fluctuations, the bias is \cite{PeaksBias}
\begin{align}
  b_h(\nu) &= 1+\frac{\nu-1}{\delta_c}
\label{kfreehbias}
\end{align}
which satisfies $\int d\nu f(\nu)b(\nu)=1$.
In detail of course the halos do not provide a linearly biased tracer of
the linear density field.  On smaller scales both higher order terms and
halo exclusion effects give rise to a scale-dependence.
Both analytic calculations \cite{PeaksBias} and simulations \cite{Sims}
suggest that this is a few percent correction on the scales of interest
to us and we shall henceforth neglect it.

When looking at large scales, such as those relevant to the baryon wiggles
in the linear power spectrum, the function $y(M,k)$ can be accurately
approximated by a Taylor expansion into powers of $kr_v$, where $r_v$ is the
virial radius, which depends upon the mass.
Assuming an NFW form \cite{NFW} for $y(M,k)$ and expressing the mass
dependence explicitly, the expression is
\begin{align}
  y(M,k)&= 1+c_2\,(kr_v)^2+c_4\,(kr_v)^4 + \cdots \nonumber \\
  &=1+c_2 (k_\star r_\star)^2 \kappa^2 m^{2/3}
     +c_4 (k_\star r_\star)^4 \kappa^4 m^{4/3} + \cdots
\end{align}
Here we have introduced another quantity, the virial radius of an $M_\star$
halo, $r_\star\equiv r_v(M_\star)= (3 M_\star/4\pi\Delta\bar{\rho})^{1/3}$,
where $\Delta$ is the virialization overdensity, which we will take
to be $\Delta=200$.  
The expansion coefficients $c_2$ and $c_4$ are functions of the halo
concentration, and for the NFW model are ratios of gamma functions.
For cluster sized halos we expect $c\simeq 5$ which leads to $c_2\simeq-0.049$
and $c_4 \simeq 0.0014$, while for galaxies $c\simeq 10$ making
$c_2\simeq -0.04$ and $c_4\simeq 0.0011$.
The quantity $k_\star r_\star$ can be computed using the relation in
Eq.~(\ref{eqn:sigdef}), and turns out to be a function only of the index $n$.
We have tabulated the expressions and values of $k_\star r_\star$ in
Table \ref{tab:nquant}.
On large scales, where $k<k_\star$, we see that the coefficients of the last
two terms in the expression for $y(M,k)$ are extremely small.
We have repeated our analysis neglecting these terms and find that these
are insignificant corrections to the scale dependence we are studying.
This is consistent with the results of \cite{SchWein}, who found that the 
local relation between galaxies and mass within a halo 
does not significantly impact
the large scale galaxy correlation function.
For simplicity we shall set $y(M,k)=1$ for the remainder of the paper.

\begin{table}
\begin{center}
\begin{tabular}{|c|c|c|c|c|c|} \hline
$n$ & \multicolumn{2}{|c|}{$k_\star r_\star$} &
  \multicolumn{2}{|c|}{$A_\star$} & $\gamma_{\rm dm}$ \\
\hline
  $0$ & $\Delta^{-1/3}\left(3 \pi/2 \delta_c^{2}\right)^{1/3}$
  & 0.2023 & $\delta_c^{-2}$ & 0.3518 & 1 \\
  $-\frac{1}{2}$ & $\Delta^{-1/3}\left(12 \sqrt{\pi}/7\delta_c^2\right)^{2/5}$ 
  & 0.1756 & $8/7 \left(12/7 \pi^2\right)^{2/5} \delta_c^{-12/5}$ & 0.2299
  & 1.18 \\
  $-1$ & $\Delta^{-1/3}\left(3/2 \delta_c\right)$
  & 0.1521 & $\left(9/4 \pi\right) \delta_c^{-3}$ & 0.1494 & 1.60 \\
  $-\frac{3}{2}$ & $\Delta^{-1/3}\left(16 \sqrt{\pi}/15\delta_c^2\right)^{2/3}$
  & 0.1303 & $\left(512/675\right)\delta_c^{-4}$ & 0.0939 & 3 \\
  $-2$ & $\Delta^{-1/3}\left(3 \pi/5\delta_c^2\right)$
  & 0.1134 & $(18 \pi^2 / 125) \delta_c^{-6}$ & 0.0619 & 15 \\ \hline
\end{tabular}
\end{center}
\caption{Expressions and values for $k_\star r_\star$ and $A_\star$ in terms
of $\delta_c=1.686$ and the virialization overdensity $\Delta=200$.
Here $k_\star$ is the normalization of the dark matter power spectrum, and 
$r_\star=r_v(M_\star)$ is the virial radius of a halo of mass $M_\star$.
The factor $A_\star=k_\star^3 M_\star/(2 \pi^2 \bar{\rho})$ relates the
amplitude of the 1- and 2-halo terms and $\gamma_{\rm dm}$ is defined in
Eq.~(\protect\ref{eqn:gdm}).}
\label{tab:nquant}
\end{table}

\section{Results -- real space}

Having argued that we can safely approximate $y$ and $b_h$ as scale independent
quantities when studying clustering at large scales our expressions simplify
dramatically.  The mass power spectrum is simply\footnote{If appropriate halo
profiles, e.g.~Gaussians, are chosen the full $k$-dependent integrals can also
be done in terms of special functions.}
\begin{equation}
  \Delta^2_{\rm dm}(k) = \kappa^{3}
  \left( \kappa^n + A_\star \gamma_{\rm dm} \right)
  \qquad \kappa \ll 1
\end{equation}
where $A_\star$ and $\gamma_{\rm dm}$ are $n$-dependent constants 
\begin{align}
  A_\star &=\frac{k_\star^3 \, M_\star}{2 \pi^2 \bar{\rho}} \\
  \gamma_{\rm dm} &= \int_0^\infty m(\nu) f(\nu) d\nu = 
    2^{3/(3+n)}\pi^{-1/2}\ \Gamma\left[1/2 + 3/(n+3)\right] \label{eqn:gdm}
\end{align}
We list the values of $A_\star$ and $\gamma_{\rm dm}$ for some values of $n$
in Table \ref{tab:nquant}.
Referring to the Table we see that, for $n$ near $-1$, the 1-halo term
dominates only for $k>k_\star$, outside of the range of relevance for us.

The scale-dependent bias can be defined as
\begin{align}\label{genb}
  B^2(k)\equiv \frac{_{\rm 2h}\Delta^2_g+_{\rm 1h}\Delta^2_g}
  {_{\rm 2h}\Delta^2_{\rm dm}+_{\rm 1h}\Delta^2_{\rm dm}}
\end{align}
which can be re-written to explicitly exhibit its scale dependence as
\begin{align}
  B^2(k)&=\left(\frac{1}{\alpha_g^2}\right)
    \frac{\beta_g^2+A_\star \gamma_g\kappa^{-n}}
    {1+A_\star \gamma_{\rm dm}\kappa^{-n}}  \label{eqn:b1} \\
  &\simeq (\beta_g/\alpha_g)^2 \, \left(1+\zeta\, \kappa^{-n} + \cdots\right)
  \label{eqn:b2}
\end{align}
where $\alpha_g$, $\beta_g$ and $\gamma_g$ are dimensionless integrals of
$\nu$, $\alpha_{g}$ is the galaxy number density in dimensionless units,
$\beta_g/\alpha_g$ is the galaxy weighted halo bias and $\gamma_{g}$ counts`
the number of galaxy pairs in a single halo\footnote{In the limit
$y(\nu,k)=1$ we have $\alpha_{\rm dm}=\beta_{\rm dm}=1$.  The expression
for $\gamma_{\rm dm}$ is given in Eq.~(\protect\ref{eqn:gdm}).}.
We have neglected terms higher order in $\kappa$.
The term $\kappa^{-n}$ encodes the leading order scale dependence and is
proportional to the inverse of the linear dark matter power spectrum.
Choosing $a=1$ in our HOD as a representative example the relevant
integrals are
\begin{align}
  \alpha_g&=\int_{\nu_{\rm min}}^\infty m(\nu)^{-1} f(\nu) 
    \left[1+m(\nu)/m_{\rm sat} \right] \,d\nu \\
  \beta_g&=\int_{\nu_{\rm min}}^\infty m(\nu)^{-1} f(\nu)
    \left[1+m(\nu)/m_{\rm sat}\right] b_h(\nu) \, d\nu \\
  \gamma_g&=\int_{\nu_{\rm min}}^\infty
    m(\nu)^{-1} f(\nu) \left[2m(\nu)/m_{\rm sat} + 
    (m(\nu)/m_{\rm sat})^2\,\right]\, \, d\nu
\label{eqn:defs}
\end{align}
and the factor governing the scale-dependence is
\begin{equation}
  \zeta(\nu_{\rm min},m_{\rm sat}, n)=
    A_\star \left(\gamma_g/\beta_g^2 - \gamma_{\rm dm} \right)
\end{equation}
Note $\zeta$ depends on the number of pairs of galaxies divided by the
square of the large-scale bias.
If one wishes to reintroduce the halo profiles there is a simple modification
to the integrals.  In $\alpha_g$, $\beta_g$, and $\gamma_g$,  every occurrence
of $m(\nu)/m_{\rm sat}$ will be multiplied by $y(\nu,k)$.
For the dark matter, $\gamma_{\rm dm}$ will have an extra factor of 
$|y(\nu,k)|^2$ in the integrand, and the $1$ in the denominator of
Eq.~(\ref{eqn:b1}) will be replaced by
$\int b_h(\nu)f(\nu)y(\nu,k) d\nu$ squared.

\begin{figure}
\begin{center}
\resizebox{4.5in}{!}{\includegraphics{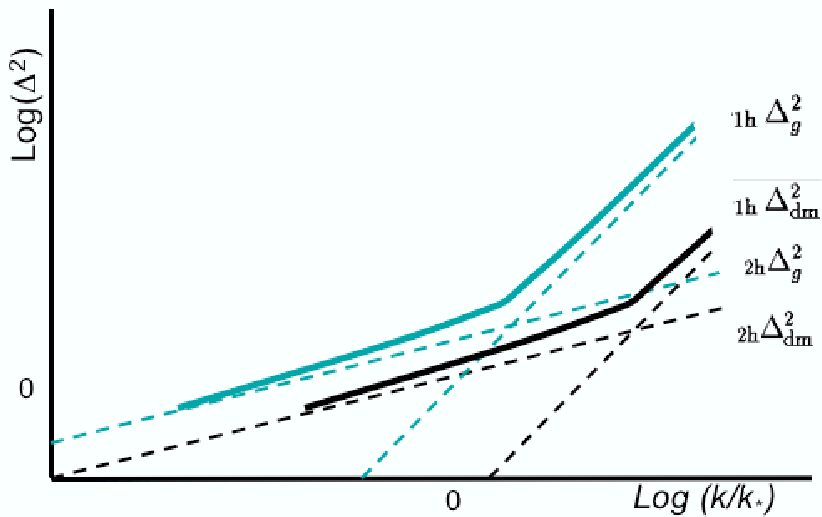}}
\end{center}
\caption{A cartoon illustrating the difference in the shift of the 
1-halo and 2-halo terms (dashed lines)
of the galaxy power spectra with respect
to the dark matter.  Because of this difference, the 1-halo term 
dominates on larger scales for the galaxy spectrum.  This leads to
a change in the ratio of the total power (solid curves), which leads
to a scale dependent galaxy bias.}
\label{fig:cartoon}
\end{figure}

In this form it is clear that the scale-dependent bias arises because
the 1- and 2-halo terms for the galaxies are different multiples of their
respective dark matter counterparts.  Typically the 1-halo term is enhanced
more than the 2-halo term, leading to an increase in the bias with decreasing
scale.  A cartoon of this is shown in Fig.~\ref{fig:cartoon}.
The relative enhancements of the 1 and 2-halo terms depend on the HOD 
parameters for the galaxy population used as tracers.
Note also that in our simple model the 2-halo term retains the oscillations
of the linear theory spectrum, while in the 1-halo term they are absent.
This provides a partial\footnote{In a more complex/realistic model the 2-halo
term involves the non-linear power spectrum and non-linear bias of the
tracers, including halo exclusion effects.  Mode coupling thus appears to
reduce the baryon signal even at the 2-halo level.} explanation of the
reduction of the contrast of the oscillations with increasing $k$.

\begin{figure}
\begin{center}
\resizebox{4.5in}{!}{\includegraphics{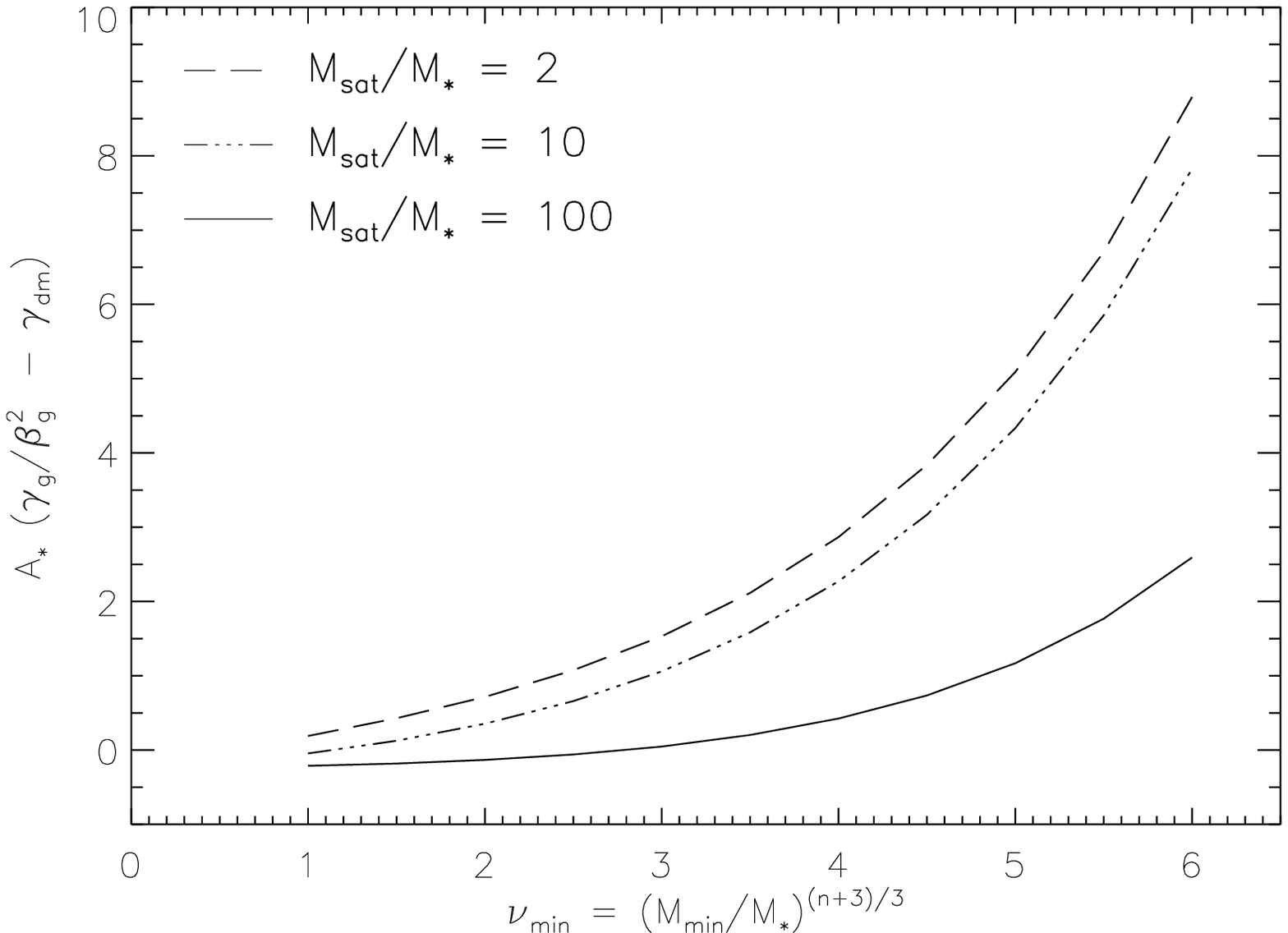}}
\end{center}
\caption{The factor $\zeta$ that governs the strength of the scale dependent 
part of the galaxy bias.}
\label{fig:zeta}
\end{figure}

Figure \ref{fig:zeta} shows $\zeta$ vs.~$\nu_{\rm min}$ for several different
values of the other HOD parameter $M_{\rm sat}$.
We see that the  scale dependence is more prominent as the number of satellite
galaxies is increased, and that a higher threshold halo mass for containing
a central galaxy leads to a more scale dependent bias at large scales.
At fixed number density $\zeta$ increases with increasing bias and is
more rapidly increasing for rarer objects.  At fixed bias $\zeta$ is larger
the rarer the object.
Our model is not sophisticated enough to expect good agreement with large
N-body simulations, however comparing to the work of \cite{Sims} we
find good qualitative agreement in the scale-dependence of the bias for
$0.01<k/(h\,{\rm Mpc}^{-1})<0.1$.

We note that the 1- and 2-halo decomposition leads us to a new
parameterization of the scale-dependent bias.
In the limit where halo profiles and scale-dependent halo bias can be
neglected the most natural description of the galaxy spectrum is
\begin{equation}
  \Delta_g^2 = b^2 \Delta_{\rm lin}^2(k) + \left( \frac{k}{k_1} \right)^3
\end{equation}
which has two free parameters, $b$ and $k_1$.  We expect this will describe
the largest part of the scale-dependent bias.  Non-linear bias, halo
exclusion and profiles will show up as smaller corrections to this formula,
such as a scale-dependence in $b$. It is difficult to compare
the scale dependence in this framework to
other treatments of scale dependent bias (e.g. \cite{soccfry}) where the galaxy
density contrast is expanded in moments of the matter 
density contrast ($b_1$, $b_2$, etc.), 
because the matter density contrast itself has both 1-halo
and 2-halo contributions, and furthermore 
we have not extended our analysis to the bispectrum
or higher order. 

\section{Results -- redshift space}

Keeping to our philosophy of examining qualitative behavior in a simple
model, we can extend these results to redshift space.
The 2-point function in redshift space differs from that in real space due
to two effects \cite{RedReview}.
The first, effective primarily on very large scales, accounts for the fact
that dark matter and the galaxies that trace it have a tendency to flow
toward overdensities as the structure in the universe is being assembled,
enhancing the fluctuations in redshift space \cite{Kaiser}.
The second comes into play inside virialized structures, where random
motions within the halo reduce fluctuations in redshift space.

These corrections impact the 1-halo and 2-halo terms.  The inflow effect
primarily impacts the 2-halo term while virial motions primarily affect
small scales which are dominated by the 1-halo term \cite{HaloRed}.
The boost in the observed density contrast, $\delta_k$, due to instreaming
is given\footnote{It has been argued in Ref.~\cite{Sco} that the form
$1+{\rm f}\mu^2$ is not highly accurate on the scales relevant to observations
and higher order corrections apply.  Since it is our intent to gain qualitative
understanding rather than quantitative accuracy we shall use the simplest
form: $1+{\rm f}\mu^2$.  Deviations from this will be yet another source of
scale-dependence, but numerical simulations suggest it is small.}
by $(1+{\rm f}\mu^2)$ where $\mu=\hat{r}\cdot\hat{k}$ and
${\rm f}\simeq\Omega_m^{0.6}$ \cite{Kaiser}.
The small scale suppression we take to be Gaussian.
In general, when extending the model to galaxies tracing the dark matter,
one should distinguish between central and satellite galaxies,
since the latter have much larger virial motions and will therefore suffer
more distortion in redshift space.  We approximate this by taking
$\sigma^2_{v,{\rm cen}}\approx 0$ and $\sigma^2_{v,{\rm sat}}=GM/2r_{\rm vir}$.
Converting from velocity to distance we have for an $M_\star$ halo
$\sigma_\star\to \sqrt{\Delta/4}\,r_\star\simeq 7\,r_\star$.
Defining $y_s(\nu,k)=y(\nu,k)\,e^{-(k\sigma_v\mu)^2/2}$ the 1- and 2-halo
terms are then given by \cite{HaloRed}
\begin{align}
  _{1h}\Delta^2_{\rm dm}
  &=\frac{k^3}{2\pi^2} \frac{M_\star}{\bar{\rho}} 
  \int_0^\infty m(\nu)\,f(\nu)\, |y_s(\nu,k)|^2\, d\nu \\
  _{2h}\Delta^2_{\rm dm}
  &=\Delta^2_{\rm lin} \left[
  \int_0^\infty f(\nu)\,(1+{\rm f}\mu^2)\,b_h(\nu)\,y_s(\nu,k)\, d\nu\right]^2
\end{align}
for the dark matter and
\begin{align}
  _{1h}\Delta^2_g &=
  \frac{k^3}{2\pi^2} \frac{\bar{\rho}}{n^2_g\,M_\star} 
  \int_{\nu_{\rm min}}^\infty m^{-1}(\nu)\, f(\nu) \nonumber \\
  &\left[
  2\left(\frac{m(\nu)}{m_{\rm sat}}\right)\,y_s(\nu,k)
  \,+\left(\frac{m(\nu)}{m_{\rm sat}}\right)^2 |y_s(\nu,k)|^2 \right] d\nu \\
  _{2h}\Delta^2_g &=
  \Delta^2_{\rm lin} \left[  \frac{\bar{\rho}}{n_g\,M_\star}
  \int_{\nu_{\rm min}}^\infty m^{-1}(\nu)\,f(\nu) \,b_h(\nu)
  \left(1+\frac{m(\nu)}{m_{\rm sat}}\,y_s(\nu,k)\right)
  \ d\nu\right. + \nonumber \\
  & \left. {\rm f}\mu^2 \int_{0}^\infty f(\nu) \,b_h(\nu)
    \, y_s(\nu,k) d\nu \right]^2
\end{align}
for the galaxies.  As discussed in \cite{HaloRed},
in the 2-halo term the effect of peculiar velocities,
going as ${\rm f}\mu^2$, is governed by the mass rather than the galaxy
density field, requiring the addition of
a separate integral over $\nu$.  Conceptually, 
this term is added to account for extra clustering in redshift 
space induced by 
the bulk peculiar flow of the galaxies in one halo under the influence of 
the dark matter in other halos.   
For some purposes it is useful to average over orientations of the
galaxy separations (i.e. integrate over $\mu$) but in the case of studying
baryon oscillations, doing so throws away valuable information.

As before we note that $y(\nu,k)\approx 1$ and
$\exp[-(k\sigma_v\mu)^2/2]\approx 1$ for $k\ll k_\star$ so the effect of
redshift space distortions is primarily to enhance the 2-halo term --
this makes the power spectrum ``more linear'' in
redshift space than real space.  However the second of our approximations,
$\exp[-(k\sigma_v\mu)^2/2]\approx 1$, is not as good as the first,
$y(\nu,k)\approx 1$, so there is enhanced $k$-dependence from the
individual terms.  For the interesting range of $n$,
$k_\star\sigma_\star\sim 1$ so the exponential can only be neglected
when  $\kappa^2\nu^{2/(n+3)}\ll 1$ for all values of $\nu$ that significantly 
contribute to the integral; i.e. near the peak of the integrand.
For example, we see scale dependence at smaller $k$ in the 1-halo term in
redshift space than in real space. At $\kappa=1/2$,
the exponential term induces a 13-14\% change in the 1-halo terms 
along the line of sight for 
both the dark matter and a moderately biased sample of galaxies,
leading to a percent level correction in the ratio
of power spectra.
The error decreases rapidly as $|\mu|$ decreases.
The importance of the exponential factor depends somewhat on the HOD parameters.
The correction to the galaxy 1-halo term 
is larger as $M_{\rm min}$ increases, but is smaller as 
$M_{\rm sat}$ increases, due to the decreasing number of satellite-satellite pairs.

For completeness we write the scale-dependent bias
in redshift space in the approximation that $y_s(\nu,k)\simeq 1$.
\begin{eqnarray}
  B^2(k,\mu)&=&\left(\frac{1}{\alpha_g^2}\right)
               \frac{\left[\beta_g+\alpha_g f \mu^2\right]^2
               + \kappa^{-n} A_\star \gamma_g}{\left[1+{\rm f}\mu^2\right]^2
               + \kappa^{-n} A_\star \gamma_{\rm dm}} \\
&\simeq& \frac{\beta_g^2}{\alpha_g^2} \,\frac{\Xi_g^2}{\Xi_{\rm dm}^2}
  \, \left( 1 + A_\star \kappa^{-n}
  \left[ \frac{\gamma_g}{\Xi_g^2} -
         \frac{\gamma_{\rm dm}}{\Xi_{\rm dm}^2} \right]
   +\cdots \right)
\end{eqnarray}
where we have defined $\Xi_g=1+(\alpha_g{\rm f}/\beta_g)\mu^2$ and
$\Xi_{\rm dm}=1+{\rm f}\mu^2$ to simplify the equations.

\section{Conclusions}

Models of structure formation where $\Omega_{\rm b}\not\ll\Omega_{\rm m}$
predict a series of features in the linear theory matter power spectrum,
akin to the acoustic peaks seen in the angular power spectrum of the cosmic
microwave background.  These peaks provide a calibrated standard ruler, and
a new route to constraining the expansion history of the universe.
In order to realize the potential of this new method, we need to understand
the conversion from what we measure -- the non-linear galaxy power spectrum
in redshift space -- to what the theory unambiguously provides -- the linear
theory matter power spectrum in real space.
The ability of N-body simulations to calibrate this mapping is improving
rapidly, but the complexity of the simulations can often mask the essential
physics.  In this paper we have tried to investigate the issues using a
simplified model which can give qualitative insights into the processes
involved.

In our toy model we find that the distribution of galaxies within halos
and the complexities of scale-dependent halo bias are sub-dominant
contributions to the scale-dependence of galaxy bias.  The dominant effect
is the relative shifts of the 1- and 2-halo terms of the galaxies compared
to the matter.  The amplitude of the scale dependent bias on very large
scales is parameterized by a quantity, $\zeta$, which depends on the
galaxy HOD.  For our two parameter HOD we find $\zeta$ increases with
increasing bias at fixed number density and is more rapidly increasing for
rarer objects.  At fixed bias, $\zeta$ is larger the rarer the object.

The 1- and 2-halo decomposition leads us to a new parameterization of the
scale-dependent bias.
In the limit where halo profiles and scale-dependent halo bias can be
neglected the most natural description of the galaxy spectrum is
\begin{equation}
  \Delta_g^2 = b^2 \Delta_{\rm lin}^2(k) + \left( \frac{k}{k_1} \right)^3
\end{equation}
which has two free parameters, $b$ and $k_1$.
This is very close to the phenomenologically motivated form proposed by
\cite{SeoEis05}.  The extra $k$-dependence these authors allowed in their
multiplicative and additive terms can be understood here as the effect
of non-linear power, non-linear bias and halo exclusion and halo profiles.
The corrections appear first in the 2-halo term and then at smaller 
scales
in the 1-halo term.
Our results also suggest that on very large scales, the bias in configuration space 
has relatively little scale dependence because the effects of the 1-halo term 
are strictly limited to scales smaller than the virial radius of the largest
halo.

We would like to thank D. Eisenstein and R. Sheth for conversations.
The simulations referred to in this work were performed on the IBM-SP
at NERSC.  This work was supported in part by NASA and the NSF.

\end{document}